\documentclass[12pt,preprint]{aastex}

\shorttitle{Twist of Magnetic Fields in Relativistic Jets}
\shortauthors{Contopoulos et al.}

\begin{document}

\title{The Invariant Twist of Magnetic Fields \\
in the Relativistic Jets of Active Galactic Nuclei}

\author{
Ioannis Contopoulos,\altaffilmark{1}
Dimitris M. Christodoulou,\altaffilmark{2} \\
Demosthenes Kazanas,\altaffilmark{3} and Denise C.
Gabuzda\altaffilmark{4} }

\altaffiltext{1}{Research Center for Astronomy, Academy of Athens,
         Athens 11527, Greece. \\
                 Email: icontop@academyofathens.gr}
\altaffiltext{2}{Dept. of Mathematical Sciences, University of
Massachusetts
Lowell, Lowell 01854. \\
                 E-mail: dimitris\_christodoulou@uml.edu}
\altaffiltext{3}{NASA/GSFC, Code 663, Greenbelt, MD 20771. \\
                 E-mail: demos.kazanas@nasa.gov}
\altaffiltext{4}{Dept. of Physics, University College Cork, Cork, Ireland. \\
                 Email: gabuzda@physics.ucc.ie}

\begin{abstract}
The origin of cosmic magnetic ({\bf B}) fields remains an open
question.  It is generally believed that very weak primordial {\bf
B} fields are amplified by dynamo processes, but it appears
unlikely that the amplification proceeds fast enough to account
for the fields presently observed in galaxies and galaxy clusters.
In an alternative scenario, cosmic {\bf B} fields are generated
near the inner edges of accretion disks in Active Galactic Nuclei
(AGNs) by azimuthal electric currents due to the difference
between the plasma electron and ion velocities that arises when
the electrons are retarded by interactions with photons.  While
dynamo processes show no preference for the polarity of the
(presumably random) seed field that they amplify, this alternative
mechanism uniquely relates the polarity of the poloidal {\bf B}
field to the angular velocity of the accretion disk, resulting in
a unique direction for the toroidal {\bf B} field induced by disk
rotation.  Observations of the toroidal fields of 29 AGN jets
revealed by parsec-scale Faraday rotation measurements show a
clear asymmetry that is consistent with this model, with the
probability that this asymmetry came about by chance being less
than 1\%. This lends support to the hypothesis that the Universe
is seeded by {\bf B} fields that are generated in AGN via this
mechanism and subsequently injected into intergalactic space  by
the jet outflows.

\end{abstract}

\section{Introduction}

Despite progress, the origin of cosmic magnetic ({\bf B}) fields
remains an open question (Carilli \& Taylor 2002; Widrow 2002;
Kronberg 2005; Kulsrud \& Zweibel 2008; Bernet et al. 2008). The
general view is that initially weak ($\sim 10^{-20}$ G) {\bf B}
fields generated by the ``Biermann battery'' (Biermann 1950) are
amplified to the observed mean galactic values ($\sim 10^{-6}$ G)
by dynamo processes. However, it is not clear how the dynamo
mechanism can achieve the observed large-scale separation of the
opposing field polarities, especially in view of flux-freezing,
which hinders field diffusion (Kulsrud \& Zweibel 2008). This
problem is particularly acute for galaxies and galaxy clusters,
which already have well-developed, ordered {\bf B} fields at
redshifts of $z \simeq 2-3$.

An alternative mechanism for the generation and amplification of
{\bf B} fields is the ``Poynting--Robertson (PR) cosmic battery,''
which has been proposed to operate in AGNs (Contopoulos \& Kazanas
1998; Contopoulos, Kazanas \& Christodoulou 2006; Christodoulou,
Contopoulos \& Kazanas 2008). In this model, radiation from the
active nucleus appears slightly anisotropic in the rest frame of
the accretion disk rotating about the central supermassive black
hole (BH) due to aberration. This gives rise to a PR drag force
that is inversely proportional to the square of the particle mass,
and so acts predominantly on the disk electrons, decreasing their
velocities relative to those of the disk protons.\footnote{From
the observer's viewpoint, this force appears when the nuclear
radiation, which is re-radiated isotropically in the rest frame of
the electrons, is beamed in the direction of motion, i.e., in the
direction of the disk rotation.} This generates azimuthal electric
currents in the direction of disk rotation, which, in turn, give
rise to a poloidal field $B_P$ whose direction is directly related
to the direction of the disk rotation. The PR current source is
much stronger than that of the Biermann battery, and furthermore
implies a direct coupling between accretion-flow vorticity,
disk-plasma diffusivity, and the generated large-scale, ordered
{\bf B} field (Contopoulos \& Kazanas 1998; Contopoulos et al.
2006; Christodoulou et al. 2008; see also Kulsrud et al. 1997 for
the role of fluid vorticity).

All that is required for this mechanism to operate is the presence
of (i) radiation from the central AGN and (ii) a surrounding
rotating accretion disk. Although Contopoulos \& Kazanas (1998)
assumed a central isotropic radiation field and an
advection-dominated accretion flow, the action of the PR battery
does not depend on the particular geometry or properties of the
AGN or accretion disk. Note also that the PR battery is a
large-scale, secular effect that is independent of (and certainly
does not preclude) the presence of small-scale {\bf B} fields and
magneto-rotational instability in the accretion disk.

In our scenario, {\bf B}-field loops generated by the PR battery
and anchored in the inner and outer accretion disk become twisted
in the azimuthal direction by the differential rotation of the
disk. As their twisting relaxes in the vertical direction, the
loops open up and separate into an ``inner'' component near the
disk symmetry axis and an ``outer'' component (the ``return
field'') threading the disk farther from the axis.  The poloidal
fields $B_{P}$ of the two components are in opposite directions,
one parallel and the other antiparallel to the angular velocity
vector ${\mathbf \omega}$.

In a general MHD picture, the direction of disk rotation and the
polarity of the ``wound-up'' {\bf B} fields are unrelated.
However, in the PR battery, the direction of $B_P$ is uniquely
determined by the direction of the disk rotation; thus, {\em the
associated jets should have a near-axis $B_P$ that is parallel to
${\mathbf \omega}$ and an extended $B_P$ with the opposite
polarity in the surrounding accretion disk, with each component
carrying the same magnetic flux.} As we will now see, this
coupling of the polarity of $B_P$ and disk rotation results in
invariant directions for the toroidal components of the inner
(near-axis) and outer (return fields), both of which are wound up
by the disk rotation.

\section{The PR Cosmic Battery and Faraday Rotation}

In the PR battery, an overall helical {\bf B} field is produced
when the bases of initially poloidal field lines are dragged by
the rotating disk plasma, with the inner $B_T$ pointing opposite
to the direction of disk rotation in the northern (N) hemisphere
of the disk and in the direction of disk rotation in the southern
(S) hemisphere (Fig.~1a); the opposite is true for the return
field farther out in the accretion disk (Fig.~1b). Fig.~1
illustrates a unique feature of the PR battery: {\em reversing the
observer's hemisphere, or equivalently, the direction of disk
rotation, reverses $B_P$, but leaves the direction of $B_T$
unchanged in the observer's sky} (red {\bf B}-field arrows in
Fig.~1).

This can be tested for AGN jets, where the direction of $B_T$ can
be directly inferred from the direction of FR gradients transverse
to the jet axis (Blandford 1993). FR is a rotation of the plane of
linear polarization that occurs when the polarized radiation
passes through a magnetized plasma. The rotation of the
polarization angle $\chi$ is determined by the observing
wavelength $\lambda$, the density of free electrons $n_e$, and the
line-of-sight component of the {\bf B} field in the plasma,
$B_{LOS}$:

\begin{equation}
\chi = \chi_0 +
 \frac{e^3\lambda^2}{8\pi \epsilon_0 m_e^2c}\int n_e(s)\vec{B}(s)\cdot\,d\vec{s}
     \ \ \equiv \ \ \chi_0 + {\rm (RM)}\lambda^2 \ ,
\end{equation}
\noindent where $\chi_0$ is the intrinsic polarization angle, $e$
the electron charge, $m_e$ the electron mass, $\epsilon_0$ the
dielectric constant, $c$ the speed of light, and RM the rotation
measure. The integral is taken over the line of sight from the
source to the observer. Upper limits to $n_e$ of a few times
$10^4\ \mbox{cm}^{-3}$ are set by the requirement that the optical
depth for free-free absorption be less than unity (assuming a
Faraday screen depth of one parsec -the same as that seen on the
plane of the sky, and a temperature of $10^4\ \mbox{K}$). RMs of
$1000\ \mbox{rad}\ \mbox{m}^{-2}$ in quasar cores to a few hundred
$\mbox{rad}\ \mbox{m}^{-2}$ for quasar jets imply fields of 3 to
0.3 $\mu\mbox{G}$ (Zavala \& Taylor~2003). The sign of the RM is
determined by the direction of $B_{LOS}$, with $B_{LOS}$ directed
toward the observer being positive. The presence of a helical jet
{\bf B} field gives rise to a transverse FR gradient, due to the
systematic change in $B_{LOS}$ across the jet. The PR battery
predicts that {\em the FR due to the inner region of helical {\bf
B} field should always increase in the CW direction on the sky,
and the FR due to the outer, return field in the CCW direction,
relative to the jet origin}.  This arises due to the unique
twisting of the inner and outer helical fields generated by the PR
battery (Fig.~1).

The direction of the net observed transverse FR gradient will be
determined by which region of helical field dominates the FR
integral. This will be determined by factors such as the fall-off
of the electron density and {\bf B} field with distance from the
jet axis and the AGN center and the opening angle and viewing
angle of the jet. If the inner (outer) helical field tends to
dominate the observed FR, this should lead to an excess of CW
(CCW) transverse FR gradients relative to the jet origins on the
corresponding scales. In contrast, if the initial poloidal field
is random, there should be equal numbers of CW and CCW gradients
within the statistical errors. Note that the FR associated with
the inner region of helical field need not be internal FR, since
it may occur predominantly in a sheath layer surrounding this
region.

\section{Faraday Rotation Data}

We searched the literature for reliable transverse FR gradients
detected across AGN jets on pc scales derived from
multi-wavelength Very Long Baseline Interferometry (VLBI)
polarimetry. We included only transverse FR gradients that are
close to orthogonal to the local jet direction, monotonic across
the jet, and extend across all or nearly all of the jet (this
excludes cases where there is an enhancement or reduction in the
local FR at one edge of the jet). This yielded the list of 29 AGN
in Table 1, which indicates the AGN, its redshift, the rough
distance from the VLBI core where the FR gradient is observed in
milliarcseconds (mas) and in pc, whether the FR gradient is CW or
CCW with respect to the jet origin, and references. FR-gradient
distances given as zero are observed across the VLBI core region
(the ``core'' corresponds to the optically thick base of the jet,
and the jet origin is further ``upstream'' from the core;
Blandford \& K\"onigl 1979). Most of these images were obtained
from 2--4~cm or 2--6~cm observations with the Very Large Baseline
Array (VLBA), and so have comparable resolutions. The only
exceptions are the 7~mm--2~cm image of 3C\,120 of Gomez et al.
(2008), 3.6--18~cm image of 1652+398 of Gabuzda (2006), and
18--22~cm image of 1749+701 of Hallahan \& Gabuzda (2009). The
case of 1803+784 is unique: while transverse FR gradients are
present in both the core region and jet, the direction of the core
gradient is constant, while the direction of the jet gradient
changes with time (Mahmud, Gabuzda \& Bezrukovs 2009). Therefore,
we have included an entry for this object only for the core-region
transverse FR gradient.

\section{Results}

Fig. 2 plots the distances from the VLBI core in pc for the
transverse FR gradients listed in Table 1, with the results
stacked vertically in order of right ascension. CW gradients are
shown as circles, and CCW gradients as triangles. For 7 AGN for
which transverse FR gradients are detected on two appreciably
different scales, the larger-scale gradients are shown by hollow
symbols.

Of the 36 total transverse FR gradients observed in these 29 AGN,
22 are CW and 14 CCW. A simple analysis using the binomial
probability distribution
\begin{equation}
P = \sum_{N_{CW}}^{N}~\frac{1}{2^{N}}~\frac{N!}{N_{CW}!N_{CCW}!} ,
\end{equation}
($N_{CW}$, $N_{CCW}$ are the numbers of CW, CCW FR gradient
measurements and $N=N_{CW}+N_{CCW}$) indicates that 22 or more CW
gradients would come about by chance with a probability of about
12\% ($N_{CW}=22, N_{CCW}=14, N=36$). However, if we consider only
the FR gradients detected closest to the VLBI core (i.e., those
depicted using filled symbols in Fig.~2), we find 22 CW gradients
and only 7 CCW gradients, with the probability of 22 or more CW
gradients coming about by chance being only 0.4\% ($N_{CW}=22,
N_{CCW}=7, N=29$)! In contrast, all 7 of the larger-scale FR
gradients (depicted by the hollow symbols in Fig.~2) are CCW. The
probability that all 7 of these 7 gradients are CCW by chance is
also low, about 0.8\% ($N_{CW}=0, N_{CCW}=7, N=7$).

These results are consistent with the expectations of the PR
battery model if the inner region of helical field dominates
relatively close to the jet origin.

\section{Discussion}

\subsection{Freedom from Observational Selection Effects}

It is important to ensure that the observed asymmetry between the
numbers of CW and CCW transverse FR gradients could not be an
observational selection effect. Our key criterion for selecting
the 29 AGN considered here is that their jets display monotonic FR
gradients that are close to orthogonal to the local jet direction.
We included all such cases, without regard to the direction of the
observed FR gradients, other than the requirement that they be
close to orthogonal to the jet. Thus, it would be essentially
impossible to inadvertently introduce a bias between the numbers
of CW and CCW gradients during our selection of the AGN for this
study.

\subsection{Potential for Deducing the Direction of Disk Rotation}

The sign of the FR observed along the central ``spine'' of the jet
is determined by the direction of the dominant poloidal field
$B_P$ and the viewing angle $\theta$.  Let us suppose that $B_P$
is generated by the PR battery and that the inner region of
helical field is usually dominant. When viewed from the N
hemisphere of the accretion disk, $B_P$ will be directed toward
the observer if $\theta$ is less than $1/\gamma$, where $\gamma$
is the Lorentz factor of the jet (i.e., the viewing angle is less
than $90^{\circ}$ in the rest frame of the jet), and away from the
observer if $\theta$ is greater than $1/\gamma$. Since the PR
battery requires that $B_P$ is parallel to ${\mathbf \omega}$ in
the inner region of the accretion disk, the direction of $B_P$
corresponds directly to the direction of rotation of the disk. If
$\theta$ is less than $1/\gamma$, positive, on average, FR values
along the jet ``spine'' indicate that the inner $B_P$ points
toward the observer, who then must be located in the N hemisphere
of the disk (as in Fig.~1a), which accordingly rotates CCW on the
sky. Similarly, if $\theta$ is less than $1/\gamma$, negative, on
average, FR values indicate that the observer is located in the S
hemisphere of the disk (as in Fig.~1b), which rotates CW on the
sky.

Unfortunately, it is not possible to be certain whether a jet's
viewing angle is less than or greater than $1/\gamma$, although
the most likely viewing angle is $\simeq 0.6\gamma$ (Kellermann et
al. 2004). For two objects in which the sense of rotation of the
underlying accretion disk is known, our model is in agreement with
the observed Faraday-rotation signs if these jets are viewed at
angles of less than $1/\gamma$. HST direct imaging of the M87
nucleus (Tsevtanov et al. 1999) and maser emission in the inner
disk of NGC 4258 (Hernstein et al. 2007) reveal that the accretion
disks in these systems rotate CW on the sky. FR measurements along
the jet spine in M87 (Zavala \& Taylor 2002) and in NGC 4258
(Krause \& Lohr 2004) yield large negative FR values ($B_{LOS}$
directed away from the observer). Thus, in the PR battery model,
an observer viewing the jet at an angle of less than $1/\gamma$
would be located in the S hemisphere (Fig.~1b), and these two
accretion disks should indeed be rotating CW on the sky.
Naturally, more such cases must be tested before it will be
possible to determine if this agreement provides further evidence
for the action of the PR battery in AGN.

\subsection{Energy Budgets of AGN}

The total magnetic energy content of the kpc-scale lobes of
extragalactic radio sources can reach $\sim 10^{60}- 10^{61}$~erg
(Kronberg et al. 2001; Carilli \& Taylor 2002; Widrow 2002;
Kronberg 2005; Kulsrud \& Zweibel 2008). If the
PR-battery-generated ${\bf B}$ fields remain near equipartition at
distances $R$ of the order of the BH event horizon, this energy is
comparable to the Poynting flux produced by the twisting of this
${\bf B}$ field ($\sim B^2R^2c$) over the AGN lifetime
($10^8-10^9$~yrs). This is also of the same order as the energy
released by the gravitational infall of matter onto the BH,
implying that AGN convert a significant fraction of their
accretion luminosity into Poynting flux, which, in turn, feeds
their radio lobes with magnetic energy.

\section{Conclusions}

We have found an asymmetry in the directions of FR gradients
observed across AGN jets that is difficult to explain using
standard models for AGN jet {\bf B} fields, in which a toroidal
field is generated when an initially random poloidal field
anchored in the accretion disk is wound up by the differential
rotation of the disk. The reason for this is simple: there should
be no preference for either the direction of the disk rotation or
the direction of the initial ("seed") poloidal field. This means
that, although the toroidal field that is generated in a
particular AGN jet will have a particular direction, there should
be no preferred direction for AGN jets as a whole.

The CW/CCW Faraday-rotation gradient asymmetry we have discovered
requires some mechanism that couples the sense of rotation of the
accretion disk to the direction of the poloidal field, and the PR
cosmic battery provides precisely such a mechanism. The observed
preference for CW FR gradients finds a natural explanation if the
poloidal fields are generated by the PR battery, and the inner,
near-axis helical {\bf B} field dominates the FR integrals in most
observed AGN jets on pc scales, with the outer field being on
occasion dominant farther from the jet origin.  The relatively few
jets that exhibit only CCW transverse FR gradients can be
understood as corresponding to the outer region of the helical
{\bf B} field, with the region of inner helical B field dominance
being on scales smaller than the resolution of the VLBI
observations. Our main conclusions can be summarized as follows:
\begin{enumerate}
\item In the PR battery model, the direction of $B_P$ threading
the inner part of the accretion disk in AGN is parallel to
${\mathbf \omega}$. The resulting FR ($B_T$) gradient near the jet
base has a unique direction projected onto the sky, increasing in
the CW direction relative to the origin of the jet, independent of
the direction of disk rotation.  We find remarkable support for
this picture: 22 of 29 transverse pc-scale FR gradients observed
closest to the VLBI cores are CW, while only 7 are CCW, with the
probability of this occurring by chance being only 0.4\%. \item In
all 7 AGN in which transverse FR gradients were observed at two
different distances from the VLBI core, the FR gradients farther
from the core are CCW, suggesting that these are regions where the
outer helical {\bf B} field dominates the total observed FR. This
also suggests the existence of a transition distance, within which
the inner region and beyond which the outer region of helical
fields tend to dominate. \item The collected pc-scale FR
observations strengthen the view that AGN jets have helical {\bf
B} fields and are primarily magnetically driven. \item Since the
total polarization rotations due to the observed FR exceed $\simeq
45^{\circ}$ ($\simeq 1$~rad) in some cases (e.g. 3C371: Gabuzda et
al. 2004; 3C78: Kharb et al. 2009), the FR in these sources must
be external (Burn 1966). This suggests that the FR due to the
inner region of helical field arises in a sheath surrounding that
region, as was also proposed by Kharb et al. (2009). FR associated
with the outer region of helical field is, of course, external.
\item If ${\bf B}$ fields are indeed produced by the PR battery in
AGN jet--disk systems and are then carried outward along jet
outflows and ultimately expelled into intergalactic space, these
fields should make a substantial contribution to the intergalactic
{\bf B} fields.
\end{enumerate}

It may be premature to state that the predominance of CW
transverse FR gradients we have found conclusively demonstrates
that the PR battery mechanism operates in the accretion disk--jet
systems of AGN. At the same time, these surprising results are
difficult to explain using any standard MHD model in the
literature, and seem to require a physical mechanism that couples
the directions of the disk rotation and the poloidal {\bf B}
field, as is the case for the PR cosmic battery.

We are currently undertaking numerical studies to investigate the
conditions required for the inner region of helical {\bf B} field
to dominate on relatively small scales, with a transition to
dominance of the outer region of helical field at some distance
along the jet. We hope that comparison of these numerical results
with observations will provide fundamentally new insight into the
inherent structures of AGN jets.

\acknowledgments We acknowledge useful discussions with Russell
Kulsrud, Ramesh Narayan, and Iossif Papadakis, and useful feedback
from an anonymous referee.

\begin{figure}[h]
\centering{
\includegraphics[angle=0,width=14.0cm]{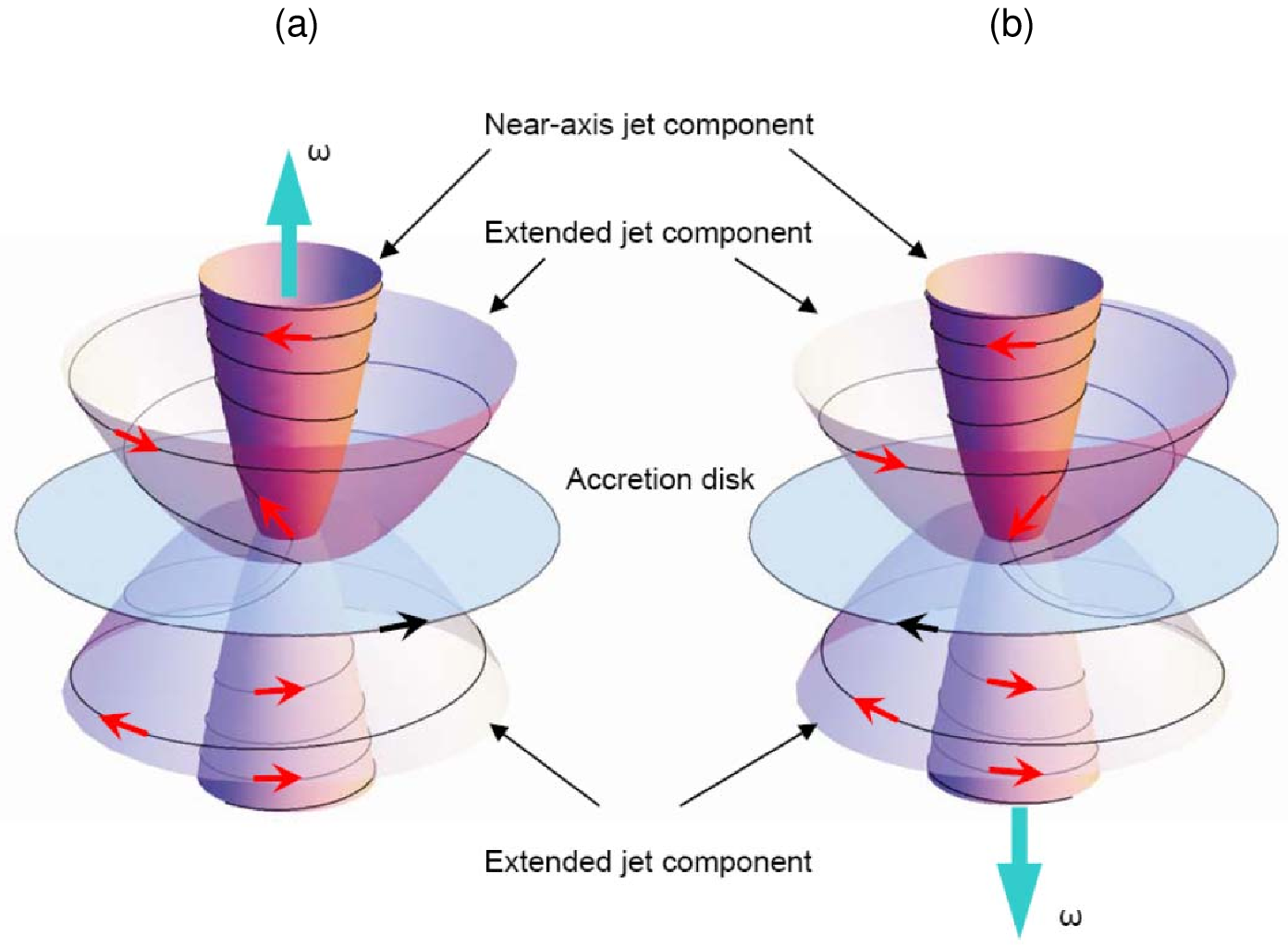}
\caption{Sketch of an AGN jet {\bf B} field generated by the PR
battery (black lines with red arrows) near the axis and periphery
of the jet.  The direction of disk rotation is shown by the black
arrows in the disk and the corresponding angular velocity vector
by the cyan arrows. The observer is located in the (a) northern
and (b) southern hemisphere of the disk.} }
\end{figure}

\begin{figure}[h]
\centering{
\includegraphics[angle=0,width=15.0cm]{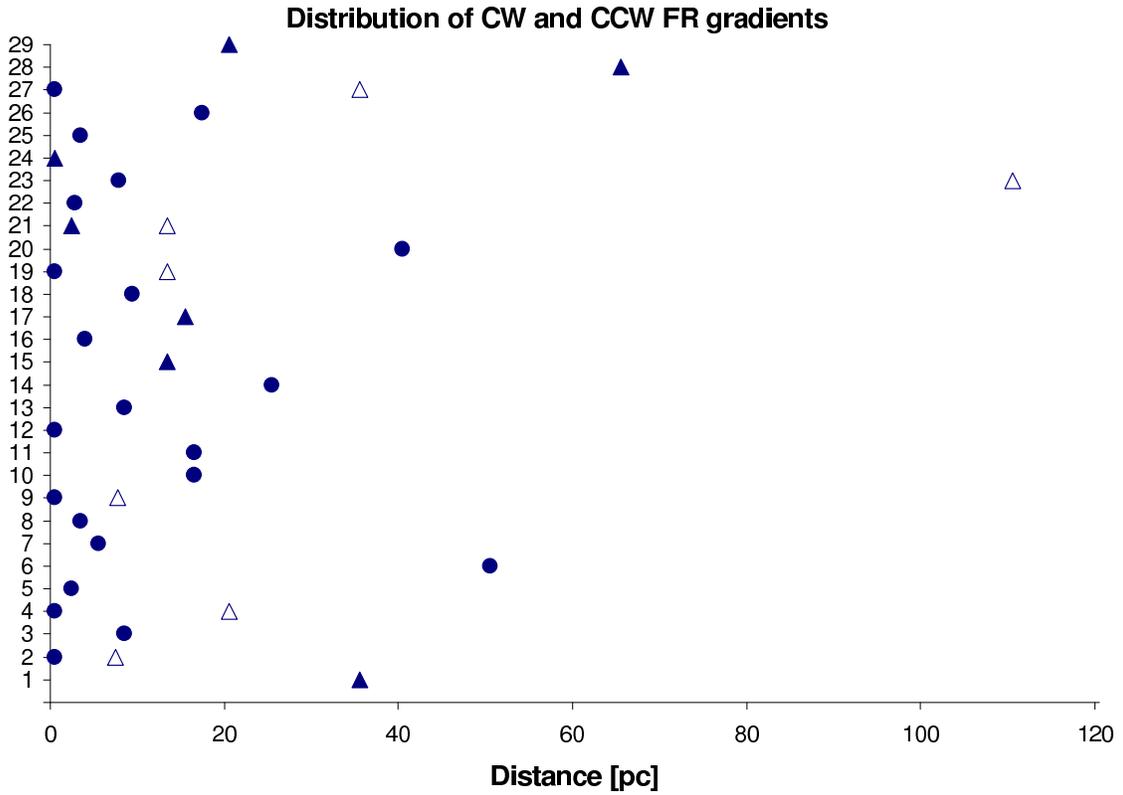}
\caption{Distribution of transverse FR gradient directions
relative to the jet origin in the plane of the sky; the numbers
correspond to those in Table 1. CW/CCW gradients are denoted with
circles/triangles, respectively. Open symbols are used for the
farther gradient in objects where FR gradients are detected on two
different scales.} }
\end{figure}

\newpage

\begin{deluxetable}{ccccccccc}
\tablecaption{AGN Jets with Transverse FR Gradient Measurements}
\tablewidth{0pt} \tablehead{ \colhead{Source} & \colhead{Object} &
\colhead{Alt.}  & \colhead{$z$} &
 \colhead{pc/mas}&\colhead{Distance} & \colhead{Distance}& Direction of
& \colhead{Refs.}\\
\colhead{Number}  &  & \colhead{Name} &  & & \colhead{(mas)} &
\colhead{(pc)}& \colhead{FR Gradient} & } \startdata
1 & 0003-066&   &  0.35& 4.91& 7& 35& CCW& 9 \\
2 & 0138-097&   &  0.733& 7.26& 0& 0& CW& 10 \\
  &     &    &      &     & 1& 7 &CCW&  \\
3 & 0212+735&   & 2.37& 8.16& 1& 8& CW& 15 \\
4 & 0256+075& &  0.90& 7.79& 0& 0& CW& 10    \\
  &     & &      &     & 2.5 & 20& CCW&  \\
5 & 0305+039 & 3C78 & 0.029 & 0.57 & 4 & 2 & CW & 19 \\
6 & 0333+321& NRAO140 & 1.259& 8.35& 6& 50& CW& 2   \\
7 & 0415+379 & 3C111&  0.05& 0.96& 5& 5& CW& 14   \\
8 & 0430+052 & 3C120&  0.03& 0.66& 4& 3& CW& 7  \\
9 & 0716+714& &  0.31& 4.56& 0& 0& CW& 10   \\
  &     & &      &     & 1.5 & 7 & CCW \\
10 & 0735+178&  DA237 & $>0.42$& $>5.4$& 3& $>16$& CW& 6 \\
11 & 0745+241& &  0.41& 5.45& 3& 16& CW& 4  \\
12 & 0748+126& & 0.889& 7.76& 0& 0& CW& 16   \\
13 & 0820+225& &  0.95& 7.9& 10& 8& CW& 4   \\
14 & 0836+710& 4C71.07 & 2.218& 8.25& 3& 25& CW& 3   \\
15 & 0954+658& &  0.37& 5.09& 2.5& 13& CCW& 9,11   \\
16 & 1156+295& 4C29.45 &  0.73& 7.26& 0.5& 3.5& CW& 6  \\
17 & 1226+023& 3C273&  0.16& 2.73& 5& 15& CCW& 1,17   \\
18 & 1253-055 & 3C279&  0.54& 6.33& 1.5& 9& CW& 16   \\
19 & 1334-127& OP158.3 &  0.54& 6.35& 0& 0& CW& 10   \\
   &    & &      &     & 2 & 13 & CCW \\
20 & 1641+399& 3C345&  0.59& 6.65& 6& 40& CW& 12  \\
21 & 1652+398 & Mrk501&  0.03& 0.66& 3& 2& CCW& 4 \\
   &     &       &      &     & 20 & 13 & CCW& 5 \\
22 & 1749+096& 4C09.57 &  0.32& 4.67& 0.5& 2.3& CW& 6   \\
23 & 1749+701& &  0.77& 7.41& 1& 7.4& CW& 8,10   \\
   &    & &      &     & 15 & 110 & CCW &  \\
24 & 1803+784 & &  0.68& 7.06& 0& 0& CCW& 15, 18 \\
25 & 1807+398 & 3C371&  0.05& 0.97& 3& 3& CW& 4 \\
26 & 2005+403 & &  1.74& 8.46& 2& 17& CW& 15  \\
27 & 2155-152 & &  0.67 & 7.03 & 0 & 0 & CW & 10 \\
   &     & &       &      & 5 & 35 &  CCW & \\
28 & 2230+114& CTA102 &  1.04& 8.08& 8& 65& CCW& 13 \\
29 & 2251+158& 3C454.3 & 0.86& 7.68& 2.5& 20& CCW& 15 \\
\enddata
\tablecomments{1 - Asada et al. 2002; 2 - Asada et al. 2008; 3 -
K. Asada 2008, private communication; 4 - Gabuzda et al. 2004; 5 -
Gabuzda 2006; 6 - Gabuzda et al. 2008; 7 - G\'omez et al. 2008; 8
- Hallahan \& Gabuzda 2009; 9 - Mahmud \& Gabuzda 2008; 10 -
Mahmud \& Gabuzda 2009; 11 - O'Sullivan \& Gabuzda 2009; 12 -
Taylor 1998; 13 - Taylor 2000; 14 - Zavala \& Taylor 2002; 15 -
Zavala \& Taylor 2003; 16 - Zavala \& Taylor 2004; 17 - Zavala \&
Taylor 2005; 18 - Mahmud et al. 2009; 19 - Kharb et al. 2009}
\label{tab:gradients}
\end{deluxetable}


\begin{thebibliography}{}
\bibitem[]{} Asada K., Inoue M., Uchida Y., Kameno S., Fujisawa K.,
Iguchi S., Mutoh M. 2002, PASJ, 54, L39
\bibitem[]{} Asada K., Inoue M., Nakamura M., Kameno S., Nagai H. 2008,
ApJ, 682, 798
\bibitem[]{} Bernet M. L., Miniati F., Lilly S. J., Kronberg P. P.,
Dassauges--Zavadsky M., Nature, 454, 302
\bibitem[]{} Biermann L. 1950, Naturforsch., 5a, 65
\bibitem[]{} Blandford R.D. 1993, in { Astrophysical Jets} (Cambridge:
Cambridge Univ. Press), p. 26
\bibitem[]{} Blandford R.D. \& K\"onigl A. 1979, ApJ, 232, 34
\bibitem[]{} Burn B.J. 1966, MNRAS, 133, 67
\bibitem[]{} Christodoulou D. M., Contopoulos I., Kazanas D. 2008, ApJ, 674, 388
\bibitem[]{} Contopoulos I. \& Kazanas D. 1998, ApJ, 508, 859
\bibitem[]{} Contopoulos I., Kazanas D., Christodoulou D. M. 2006,
ApJ, 652, 1451
\bibitem[]{} Carilli C. L. \& Taylor G. B. 2002, Ann. Rev. Astron. Astrophys.,
40, 319
\bibitem[]{} Gabuzda D. 2006, in { Proceedings of the 8th European VLBI
Network Symposium},
http://pos.sissa.it//archive/conferences/036/011/8thEVN\_011.pdf
\bibitem[]{} Gabuzda D.~C., Murray \'E., Cronin P. 2004, MNRAS, 351, L89
\bibitem[]{} Gabuzda D. C., Vitrishchak V. M., Mahmud M., O'Sullivan S. P.
2008, MNRAS, 384, 1003
\bibitem[]{} G\'omez J. L., Marscher A. P., Jorstad S. G., Agudo I.,
Roca--Sogorb M. 2008, ApJ, 681, L69
\bibitem[]{} Hallahan R. \& Gabuzda D. C. 2009,  in { Proceedings of the 9th
EVN Symposium}, in press (Proceedings of Science, 2009)
\bibitem[]{} Hernstein, J. R., Moran, J. M., Greenhill, L. J.
\& Trotter, A. S. 2007, ApJ, 629, 719
\bibitem[]{} Kellermann K.I., Lister M.L., Homan D.C., Vermeulen R.C.,
Cohen M.H., Ros E., Kadler M, Zensus J.A. \& Kovalev Y.Y. 2004,
ApJ, 609, 539
\bibitem[]{} Kharb P., Gabuzda D.C., O'Dea C.P., Shastri P. \& Baum S.A.
2009, ApJ, 694, 1485
\bibitem[]{} Krause, M. \& Lohr, A. 2004, A\& A, 420, 115
\bibitem[]{} Kronberg P. P. 2005, in { Cosmic Magnetic Fields}, Lecture
notes in Physics, R. Wielebinski and R. Beck (Eds.), 664, 9
(Springer: Berlin  2005)
\bibitem[]{} Kronberg P. P., Dufton Q. W., Li H., Colgate S. A. 2001, ApJ,
560, 178
\bibitem[]{} Kulsrud R. M., Cen R., Ostriker J. P., Ryu D. 1997, ApJ,
480, 481
\bibitem[]{} Kulsrud R. M \& Zweibel E. G. 2008, Rep.
Prog. Phys., 71, 046901, doi:10.1088/0034-4885/71/4/046901
\bibitem[]{} Mahmud M. \& Gabuzda D. 2008, in { Extragalactic Jets: Theory
and Observations from Radio to Gamma Ray}, ASP Conf. Ser., 386,
494
\bibitem[]{} Mahmud M. \& Gabuzda D. C. 2009,  in { Proceedings of the 9th
EVN Symposium} (Proceedings of Science, 2009);
http://pos.sissa.it//archive/conferences/072/011/IX\%20EVN\%20Symposium\_011.pdf
\bibitem[]{} Mahmud M., Gabuzda D. C. \& Bezrukovs V. 2009, MNRAS, in press;
arXiv:astro-ph/0905.2368
\bibitem[]{} O'Sullivan S.P. \& Gabuzda D. C. 2009, MNRAS, 393, 429
\bibitem[]{} Taylor G. B. 1998, ApJ, 506, 637
\bibitem[]{} Taylor G. B. 2000, ApJ, 533, 95
\bibitem[]{} Tsvetanov, Z. I., Allen, M. G., Ford, H. C.
\& Harms, R. J., in { The Radio Galaxy Messier 87}, H.-J.
R\"{o}ser \& K. Meisenheimer (eds.), 301 (Berlin: Springer 1999)
\bibitem[]{} Widrow L. M., Rev. Mod. Phys., 74, 775
\bibitem[]{} Zavala R. T. \& Taylor G. B. 2002, ApJ, 566, L9
\bibitem[]{} Zavala R. T. \& Taylor G. B. 2003, ApJ, 589, 126
\bibitem[]{} Zavala R. T. \& Taylor G. B. 2004, ApJ, 612, 749
\bibitem[]{} Zavala R. T. \& Taylor G. B. 2005, ApJ, 625, L73
\end{thebibliography}
\end{document}